\documentclass[reprint,amsmath,amssymb,aps,superscriptaddress]{revtex4-2}
\pdfoutput=1
\usepackage{graphicx}
\usepackage{dcolumn}
\usepackage{bm}

\begin{document}

\preprint{APS/123-QED}

\title{Fluctuations of Heat in Driven Two-State Systems: Application to Single-Electron Box with Superconducting Gap}

\author{Tuomas Pyh{\"a}ranta}
\affiliation{Pico group, QTF Centre of Excellence, Department of Applied Physics, Aalto University, P.O. Box 15100, FI-00076 Aalto, Finland
}
\author{Luca Peliti}
\affiliation{Santa Marinella Research Institute, Santa Marinella, Italy}
\author{Jukka P. Pekola}
\affiliation{Pico group, QTF Centre of Excellence, Department of Applied Physics, Aalto University, P.O. Box 15100, FI-00076 Aalto, Finland
}

\date{\today}

\begin{abstract}
We use trajectory averaging to show that the energy dissipated in the nonequilibrium energy-state transitions of a driven two-state system satisfies a fluctuation-dissipation relation. 
This connection between the average energy dissipation induced by external driving and its fluctuations about equilibrium is preserved by an adiabatic approximation scheme. 
We use this scheme to obtain the heat statistics of a single-electron box with superconducting leads in the slow-driving regime, where the dissipated heat becomes normally distributed with a relatively high probability to be extracted from the environment rather than dissipated. 
We also discuss heat fluctuation relations for driven two-state transitions. 

\end{abstract}

\maketitle

As the methods of fabricating, operating, and controlling electronic systems reach smaller and smaller scales, it is becoming increasingly important to understand the influence of fluctuations on different properties of these systems. 
Various fluctuation relations have been proposed \cite{BochkovKuzozlev,Jarzynski, Crooks1, Crooks2, Seifert} to describe the inherent asymmetry between the probabilities of observing specific events of fluctuating quantities and their time-reversed counterparts \cite{PelitiPigolotti}. 
This asymmetry is a signature of irreversible processes, which are accompanied by the production of entropy and the dissipation of energy. 
Here, we explore the statistics of energy dissipation in driven two-state transitions. 
While our studies are general and apply to all systems whose operation can be restricted to two states, which have proven useful in the verification of fluctuation relations \cite{tls1, tls2}, experiments \cite{exp1, exp2, exp3, exp5} have shown that single-electron devices, which enable the precise control of charge, make for an ideal platform for the testing of fluctuation relations in electronic systems. 
For this reason, we use the single-electron box \cite{seb1, seb2}, which has been used in a number of studies \cite{exp1, exp2, exp5, seb5, seb6, AverinLikharev} on fluctuations owing to the relative simplicity of its modeling, fabrication, and operation, as an example to illustrate our results. 
The single-electron box comprises a metallic island coupled to a superconducting electrode through a normal metal-insulator-superconductor (NIS) tunnel junction, and a gate voltage biases the island with respect to the electrode, creating a chemical-potential difference between them. 
At low temperatures, the dynamics of the box are restricted to just two states: no electrons on the island and one electron on the island. 
This setup is particularly interesting to consider as the use of a superconducting material for the leads opens a gap in the density of states \cite{bcs}, and the gate voltage can be dynamically varied to sweep the energy bias between the island and the electrode back-and-forth across the gap. 
This superconducting gap influences the rate of tunneling between the lead and the island, and owing to its finite width we need to drive the gate voltage slowly to accurately capture the effects of subgap tunneling on the energy-dissipation statistics. 
Since the system remains close to equilibrium under adiabatic driving, it is possible to analytically evaluate the thermodynamic quantities of interest, which are in general only treatable by numerical methods.

\textit{Master equation.---}
We model the two-state dynamics classically within a Markovian framework. 
The occupation probabilities $P_0$ and $P_1$ for energy states $n=0$ and $n=1$, respectively, evolve in time according to a linear master equation, which can be reduced to a single equation $\dot{P} = -\Gamma_\Sigma P + \Gamma_\mathrm{D}$ for the difference $P \equiv P_1 - P_0$ with the steady-state solution $\bar{P} = \Gamma_\mathrm{D}/\Gamma_\Sigma$. 
Here, $\Gamma_\Sigma \equiv \Gamma_+ + \Gamma_-$ and $\Gamma_\mathrm{D} \equiv \Gamma_+ - \Gamma_-$ are defined in terms of the transition rates $\Gamma_+$ and $\Gamma_-$ from state $0$ to state $1$ and vice versa, respectively. 
We assume that these rates satisfy the detailed-balance condition $\Gamma_+/\Gamma_- = \mathrm{e}^{\beta\mu}$ with $\beta \equiv 1/k_\mathrm{B}T$ denoting the inverse temperature and $\mu$ the energy difference between the two states. 
From the solution to the master equation subject to a general initial-value condition $P(t) = \int_{t_0}^t \mathrm{d}\tau_1 \, \Gamma_\mathrm{D}(\tau_1) \, \mathrm{e}^{-\int_{\tau_1}^t \mathrm{d}\tau_2 \, \Gamma_\Sigma(\tau_2)} + p(t_0) \, \mathrm{e}^{-\int_{t_0}^t \mathrm{d}\tau \, \Gamma_\Sigma(\tau)}$, found using an integrating factor, we can use the identity $P_n=(1-(-1)^n P)/2$ to extract the conditional transition probabilities of the system being in state $n_2$ at time $t_2$ provided it was in state $n_1$ at time $t_1 < t_2$. 
These are given jointly by the expression \cite{Averin2011} 
\begin{align}
    P_{n_1 n_2}(t_1,t_2) =& \frac{1}{2} \Big[ 1 + (-1)^{n_1+n_2} \, \mathrm{e}^{-\int_{t_1}^{t_2} \mathrm{d}\tau \, \Gamma_\Sigma(\tau)} \label{eq:Conditional} \\
    &- (-1)^{n_2} \int_{t_1}^{t_2} \mathrm{d}\tau_1 \, \Gamma_\mathrm{D}(\tau_1) \, \mathrm{e}^{-\int_{\tau_1}^{t_2} \mathrm{d}\tau_2 \, \Gamma_\Sigma(\tau_2)} \Big]. \nonumber 
\end{align}
From now on, we shall assume that the system is initially prepared in the state $n=0$; that is, $P_{n}(t) = P_{0n}(-\infty,t)$.

\textit{Fluctuation-dissipation relation.---}
Having established the dynamical framework, we can now explore the thermodynamics of driven two-state transition events. 
We are specifically interested in the statistics of the energy that is dissipated during a single ramp of the energy bias $\mu$ between the energy states, which can be obtained using trajectory averaging. 
To gain analytical results, we assume that the energy bias can be linearized as $\mu = \dot{\mu}t$ with $\dot{\mu}$ a constant on an experimentally relevant energy scale, though we emphasize that our results can be extended to nonlinear driving with minor adjustments. 
Furthermore, the linear ramp is representative of the examples considered here, where transitions occur in a small interval around the degeneracy point $\mu=0$. 
The heat generated during a single ramp is a stochastic variable comprising contributions from a random number of jumps at random times between the energy states, averaging over an ensemble of ramps to 
\begin{align}
    \langle Q \rangle = \int_{-\infty}^\infty \mathrm{d}t \, \mu\langle\dot{n}\rangle = \int_{-\infty}^\infty \mathrm{d}t \, \mu\dot{P}_1. 
\end{align}
To ease our analysis, we have taken the integration boundaries to infinity, which is justified as transitions between different states must be suppressed at large absolute values of the linearly time-dependent energy bias. 
The full solution to the master equation is generally difficult to treat analytically, so we expand the probability $P_1 \equiv \bar{P}_1 + \delta P_1$ around the quasistatic solution $\bar{P}_1 = \Gamma_+/\Gamma_\Sigma$. 
The correction $\delta P_1$, accounting for the effects of nonquasistatic driving, is exact within our Markovian model. 
The advantage of this expansion becomes apparent when considering the contributions of the two terms to the mean dissipated energy: for the first term, the energy dissipated at negative energy biases is canceled out at positive ones, leaving the mean fully determined by the correction. 
Substituting the expansion into the master equation $\dot{P}_1 = \Gamma_+ P_0 - \Gamma_- P_1$, we find $\delta \dot{P}_1 = -\Gamma_\Sigma \delta P_1 - \dot{\bar{P}}_1 = -\Gamma_\Sigma \delta P_1 - \beta\dot{\mu}\bar{P}_0\bar{P}_1$ or 
\begin{align}
    \delta P_1(t) = - \beta\dot{\mu} \int_{-\infty}^t \mathrm{d}\tau_1 \, \bar{P}_0(\tau_1) \bar{P}_1(\tau_1) \, \mathrm{e}^{-\int_{\tau_1}^t \mathrm{d}\tau_2 \, \Gamma_\Sigma(\tau_2)} \label{eq:FullCorrection}
\end{align}
when the system begins and ends each ramp in the definite states $n=0$ and $n=1$, respectively. 
Following an integration by parts for $\langle Q\rangle = -\int_{-\infty}^\infty \mathrm{d}t \, \dot{\mu} \, \delta P_1$, we find the energy associated with the correction term 
\begin{align}
    \langle Q \rangle = \beta\dot{\mu}^2\int_{-\infty}^\infty \mathrm{d}t \int_{-\infty}^t \mathrm{d}\tau_1 \, \bar{P}_0(\tau_1) \bar{P}_1(\tau_1) \, \mathrm{e}^{-\int_{\tau_1}^t \mathrm{d}\tau_2 \, \Gamma_\Sigma(\tau_2)}. \label{eq:Mean}
\end{align}
Similarly, we can use the jump-trajectory method to calculate the fluctuations of dissipated energy $\langle\delta Q^2\rangle \equiv \langle (Q-\langle Q\rangle)^2\rangle$ according to 
\begin{align}
    \langle\delta Q^2\rangle = \dot{\mu}^2 \int_{-\infty}^\infty \mathrm{d} t_1 \int_{-\infty}^\infty \mathrm{d} t_2 \, g(t_1,t_2), \label{eq:Variance}
\end{align}
where we have introduced the correlation function \cite{Korotkov} 
\begin{align}
\begin{split}
    g(t_1,t_2) = \sum_{n_1 n_2} & P_{n_1}(t_<) P_{n_1 n_2}(t_<,t_>) \\
    &\times \big(n_1-\langle n(t_<)\rangle\big)\big(n_2-\langle n(t_>)\rangle\big) 
\end{split}
\end{align}
with $t_< \equiv \min\{t_1,t_2\}$, $t_> \equiv \max\{t_1,t_2\}$. 
After a little algebra, we find $g(t_1,t_2) = P_0(t_<) P_1(t_<) (P_1(t_<) - P_{01}(t_<,t_>)) = P_0(t_<) P_1(t_>) \, \mathrm{e}^{-\int_{t_<}^{t_>} \mathrm{d}\tau \, \Gamma_\Sigma(\tau)}$. 
To obtain the last equality, we have assumed that the improper integral $\int_{-\infty}^t \mathrm{d}\tau \, \Gamma_\Sigma(\tau)$ from Eq.~\eqref{eq:Conditional} diverges. 
Finally, as fluctuations induced by slow driving are accurately described by the quasistatic solutions $\bar{P}_0$ and $\bar{P}_1$ to the master equation, 
\begin{align}
    \langle\delta Q^2\rangle \simeq 2\dot{\mu}^2 \int_{-\infty}^\infty \mathrm{d}t_2 \int_{-\infty}^{t_2} \mathrm{d}t_1 \, \bar{P}_0(t_1) \bar{P}_1(t_1) \, \mathrm{e}^{-\int_{t_1}^{t_2} \mathrm{d}\tau \, \Gamma_\Sigma(\tau)}, \label{eq:Noise}
\end{align}
where we have exploited the time-symmetry property of the correlation function $g(t_1,t_2) = g(t_2,t_1)$. 
Comparing Eqs. \eqref{eq:Mean} and \eqref{eq:Noise}, we observe that 
\begin{align}
    \langle\delta Q^2\rangle = 2 k_\mathrm{B}T \langle Q\rangle. \label{eq:FluctuationDissipationRelation}
\end{align}
This fluctuation-dissipation relation links the quasiequilibrium fluctuations of the dissipated energy with the mean dissipation induced by the slow driving perturbing the system out of equilibrium. 
It should be noted that if the distribution of heat is Gaussian, then Eq.~\eqref{eq:FluctuationDissipationRelation} is implied by the integral fluctuation relation \cite{PelitiPigolotti}.

\textit{Adiabatic approximation.---}
The integrals of Eqs. \eqref{eq:Mean} and \eqref{eq:Noise}, while exact, do not generally admit analytical expressions even for simple experimental setups, as we shall later demonstrate. 
Nonetheless, analytical estimates for the statistics of dissipated energy can be found under slow driving, where we can calculate an adiabatic correction to the mean heat. 
To this end, we assume that the system response to the changing energy bias satisfies $\delta\dot{P}_1/\dot{P}_1 \ll 1$, which reduces the master equation to a simple algebraic equation for the adiabatic correction: 
\begin{align}
    \delta P_1 \simeq -\dot{\bar{P}}_1/\Gamma_\Sigma = -\beta\dot{\mu}\bar{P}_0\bar{P}_1/\Gamma_\Sigma. \label{eq:AdiabaticCorrection} 
\end{align}
Substituting this into $\langle Q\rangle = -\int_{-\infty}^\infty \mathrm{d}t \, \dot{\mu} \, \delta P_1$ yields an adiabatic estimate for the mean. 
Next, to implement a similar approximation for variance, it becomes convenient to reverse the order of integration in Eq.~\eqref{eq:Noise}: $\langle\delta Q^2\rangle \simeq \int_{-\infty}^\infty \mathrm{d}t_1 \, \bar{P}_0(t_1) \bar{P}_1(t_1) \int_{t_1}^\infty \mathrm{d}t_2 \, \mathrm{e}^{-\int_{t_1}^{t_2} \mathrm{d}\tau \, \Gamma_\Sigma(\tau)}$. 
If the rates vary slowly at small energy biases, we can apply the linear approximation $\int_{t_1}^{t_2} \mathrm{d}\tau \, \Gamma_\Sigma(\tau) \simeq \Gamma_\Sigma(t_1)(t_2-t_1)$. 
This approximation is justified, because the double integral in $\langle\delta Q^2\rangle$ is dominated by small values of $t_1$ and $t_2-t_1$. 
In the end, we are left with 
\begin{align}
    \langle\delta Q^2\rangle &\simeq 2\dot{\mu}^2 \int_{-\infty}^\infty \mathrm{d}t \, \bar{P}_0(t) \bar{P}_1(t)/\Gamma_\Sigma(t). \label{eq:AdiabaticNoise}
\end{align}
Comparing the adiabatic correction in Eq.~\eqref{eq:AdiabaticCorrection} and the integrand of \eqref{eq:AdiabaticNoise}, we observe that our analytical estimates also satisfy the fluctuation-dissipation relation \eqref{eq:FluctuationDissipationRelation} as a sign of consistency between exact results and our approximation scheme in the limit of adiabatic driving, where the fluctuation-dissipation relation from Eq.~\eqref{eq:FluctuationDissipationRelation} can be expected to hold.

\begin{figure}[t]
    \centering
    \includegraphics[width=\linewidth]{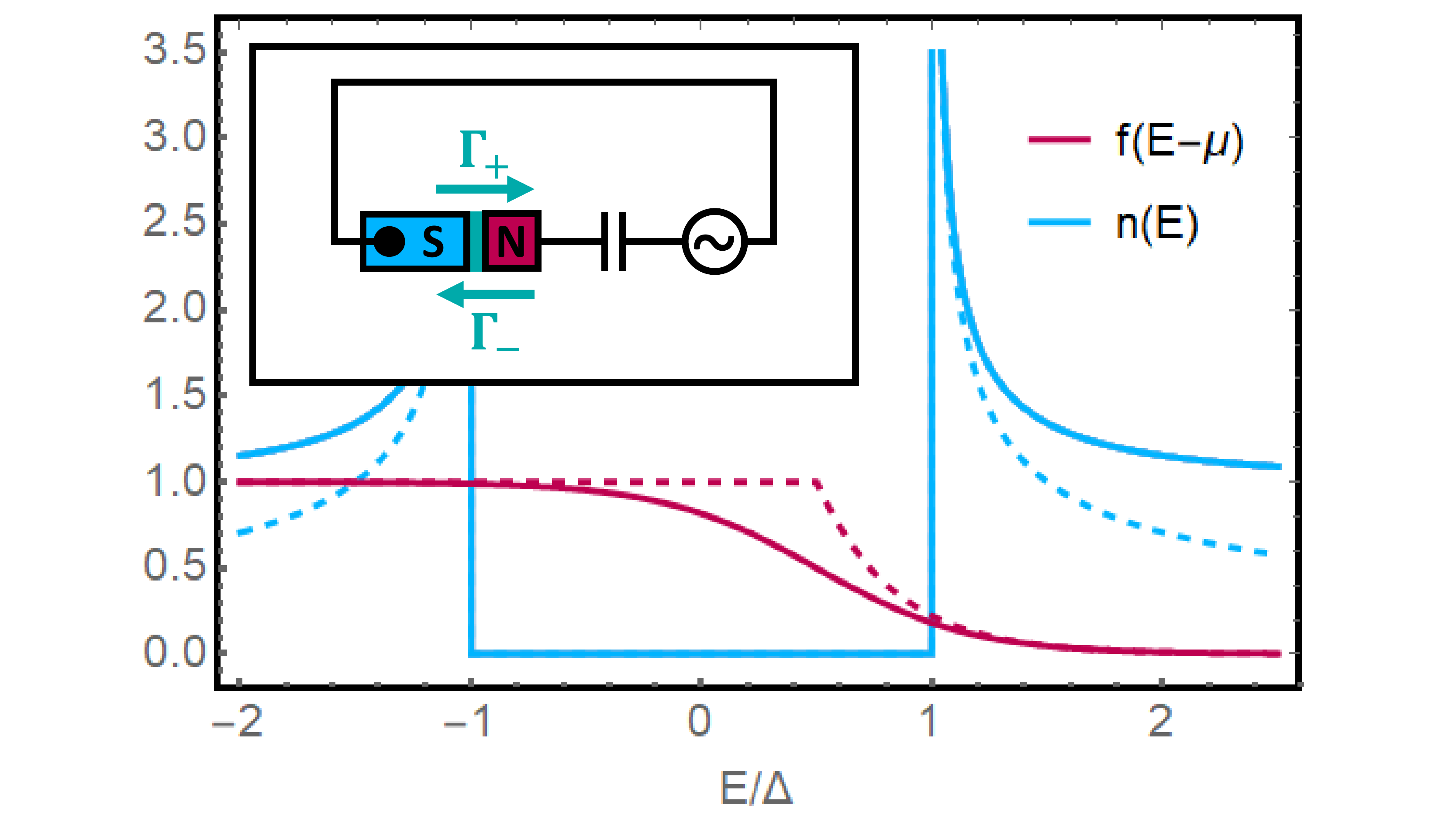}
    \caption{
    Approximations for calculating NIS tunneling rates. 
    As outlined in the main text, the errors resulting from our approximations, represented in the figure with dashed lines, for the Fermi-Dirac distribution and the BCS density of states, shown here in solid red and blue lines, respectively, are concentrated inside the superconducting gap and far outside its edges. 
    Here, we have chosen $\beta\Delta = 3$ and $\mu/\Delta = 1/2$ for demonstrative purposes; using more realistic temperatures, e.g., $\beta\Delta \gtrsim 10$ for superconducting aluminum, the presented approximations work extremely well. 
    The inset schematically depicts the single-electron box with superconducting leads. 
    Tunneling between the superconducting electrode and the normal-state metallic island and vice versa happens across the insulating barrier at rates $\Gamma_+$ and $\Gamma_-$, respectively. 
    }
    \label{fig:NIS_error}
\end{figure}
\textit{Example: single-electron box.---}
As a demonstration of our results, we apply our analysis to a single-electron box, which can be modeled as a two-state system at low temperatures. 
With each electrode at the same temperature, electrons tunnel across an NIS tunnel junction at rates given by the golden-rule expressions 
\begin{align}
\begin{split}
    \Gamma_+ =& \frac{G_\mathrm{T}}{e^2} \int_{-\infty}^\infty \mathrm{d}E \, n(E) f(E-\mu) \big( 1 - f(E) \big), \\
    \Gamma_- =& \frac{G_\mathrm{T}}{e^2} \int_{-\infty}^\infty \mathrm{d}E \, n(E) f(E) \big( 1 - f(E-\mu) \big) \label{eq:TunnelingRates}
\end{split}
\end{align}
with $n(E)=\left|\mathrm{Re}\big\{E/\sqrt{E^2-\Delta^2}\big\}\right|$ denoting the electrode density of states given by BCS theory and $f(E) = 1/(\mathrm{e}^{\beta E}+1)$ the Fermi-Dirac distribution. 
To gain analytical expressions for the tunneling rates, we first use particle-hole symmetry $1-f(E) = f(-E)$ to write the tunneling rates in the more compact form 
\begin{align}
    \Gamma_\pm = \frac{G_\mathrm{T}}{e^2} \int_{-\infty}^\infty \mathrm{d}E \, n(E) f(\pm(E-\mu)) f(\mp E). \label{eq:TunnelingRatesSymmetry}
\end{align}
Next, we approximate the tail of the Fermi-Dirac distribution with its asymptotic behavior $f(E) \sim \mathrm{e}^{-\beta E}$ while treating the states below the energy bias as filled; that is, $f(E-\mu) \simeq \mathrm{e}^{-\beta(E-\mu)}$ and $f(E-\mu) \simeq 1$ for $E > \mu$ and $E \le \mu$, respectively. 
We also expand the density of states as $n(E) \simeq 1/\sqrt{2(\pm E/\Delta - 1)}$ just outside the gap of the superconductor. 
To justify these approximations, we note that at low temperatures the resulting errors, depicted graphically in Fig.~\ref{fig:NIS_error}, are concentrated at energies with little influence on the tunneling-rate integrals in Eq.~\eqref{eq:TunnelingRatesSymmetry}. 
With small energy biases, the error due to our approximation for $f(E-\mu)$ is confined to the superconducting gap at low temperatures, where the Fermi-Dirac distribution decays rapidly above the energy bias, and the gap does not contribute to the tunneling rates as the density of states vanishes. 
Conversely, our density-of-states approximation only begins to deviate from the BCS result far outside the superconducting gap, where the decay of the Fermi-Dirac distribution prevents errors from accumulating in the tunneling rates at small biases. 
While the errors from our approximations may be larger further away from the degeneracy, the thermodynamic behavior of our two-state system is dominated by small values of the energy bias owing to the suppression of single-electron transitions at larger biases. 
Evaluating the integrals in Eq.~\eqref{eq:TunnelingRatesSymmetry}, we find the tunneling rates 
\begin{align}
\begin{split}
    \Gamma_\pm &\simeq \frac{G_\mathrm{T}\Delta}{e^2} \sqrt{\frac{\pi}{2\beta\Delta}} \, \mathrm{e}^{-\beta\Delta} (1+\mathrm{e}^{\pm\beta\mu}) \label{eq:NISTunnelingRates} 
\end{split}
\end{align}
inside the gap and near its edges. 
We now deploy the adiabatic approximation, and to this end it becomes convenient to define a dimensionless driving rate 
\begin{align}
    \lambda \equiv \frac{\beta\dot{\mu}}{\Gamma_\pm(\mu=0)} = \frac{\beta\dot{\mu}e^2}{G_\mathrm{T} \Delta} \sqrt{\frac{\beta\Delta}{2\pi}} \, \mathrm{e}^{\beta\Delta} 
\end{align}
to quantify the adiabaticity of the drive. 
Ideally, as the time-scales for external driving and tunneling are set by $\beta\dot{\mu}$ and $\Gamma_\pm(\mu=0)$, respectively, we want $\lambda \ll 1$ for the adiabatic approximation to give accurate results, and hence the ramping rate $\dot{\mu}$ must be decreased superexponentially as temperature is lowered to maintain adiabatic driving. 
With this condition satisfied, the mean and fluctuations are connected through the fluctuation-dissipation relation 
\begin{figure}[t]
    \centering
    \includegraphics[width=\linewidth]{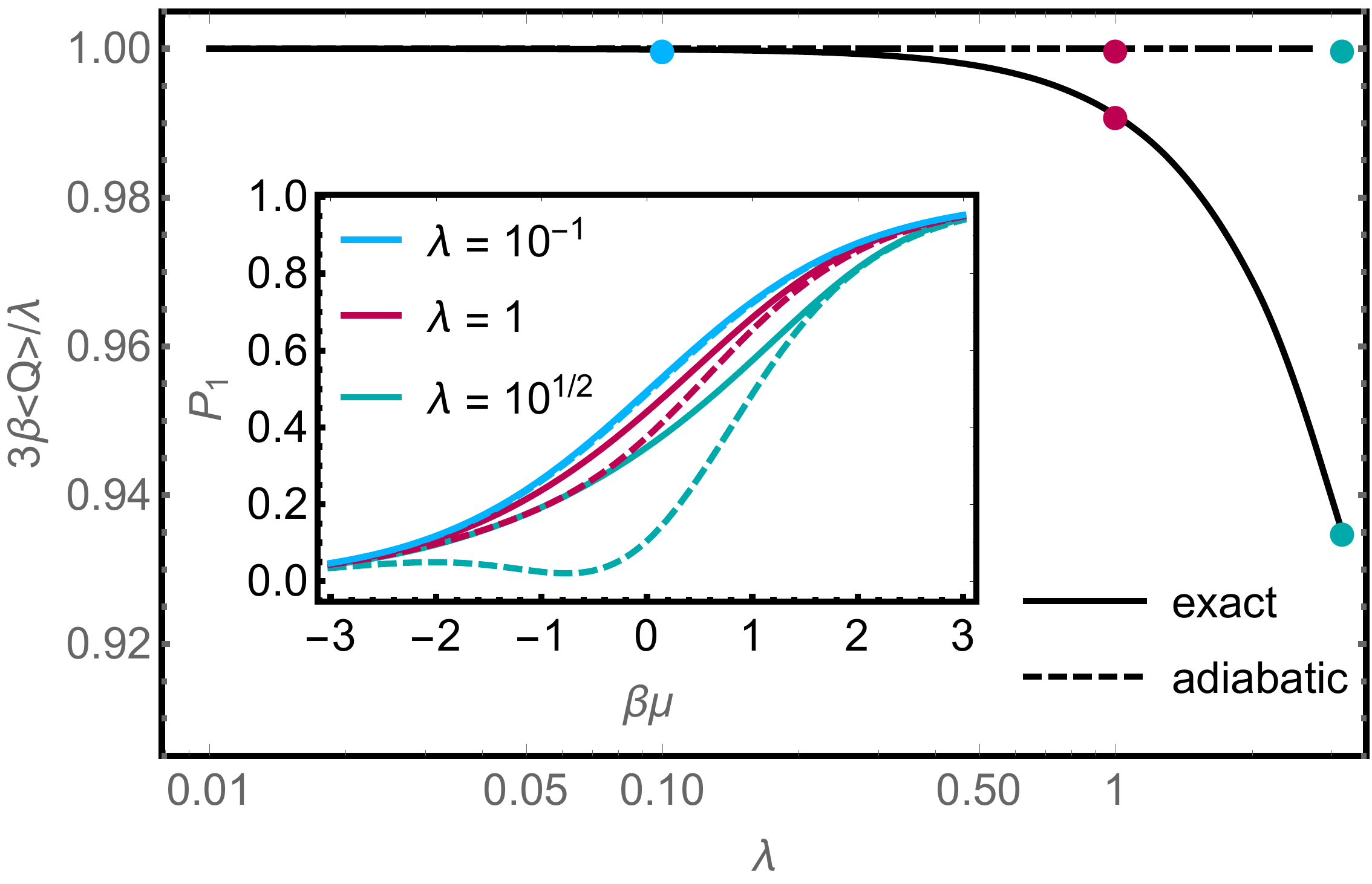}
    \caption{
    Mean dissipated energy $\langle Q \rangle$ scaled by its adiabatic estimate against the dimensionless driving rate $\lambda$. 
    The inset shows the influence of driving rate on the accuracy of the adiabatic approximation with the evolution of $P_1$ at low, intermediate, and high driving rates depicted here in blue, red, and green, respectively. 
    Solid lines represent exact results and dashed lines the adiabatic approximation in the main figure as well as the inset. 
    }
    \label{fig:NIS_variances}
    \includegraphics[width=\linewidth]{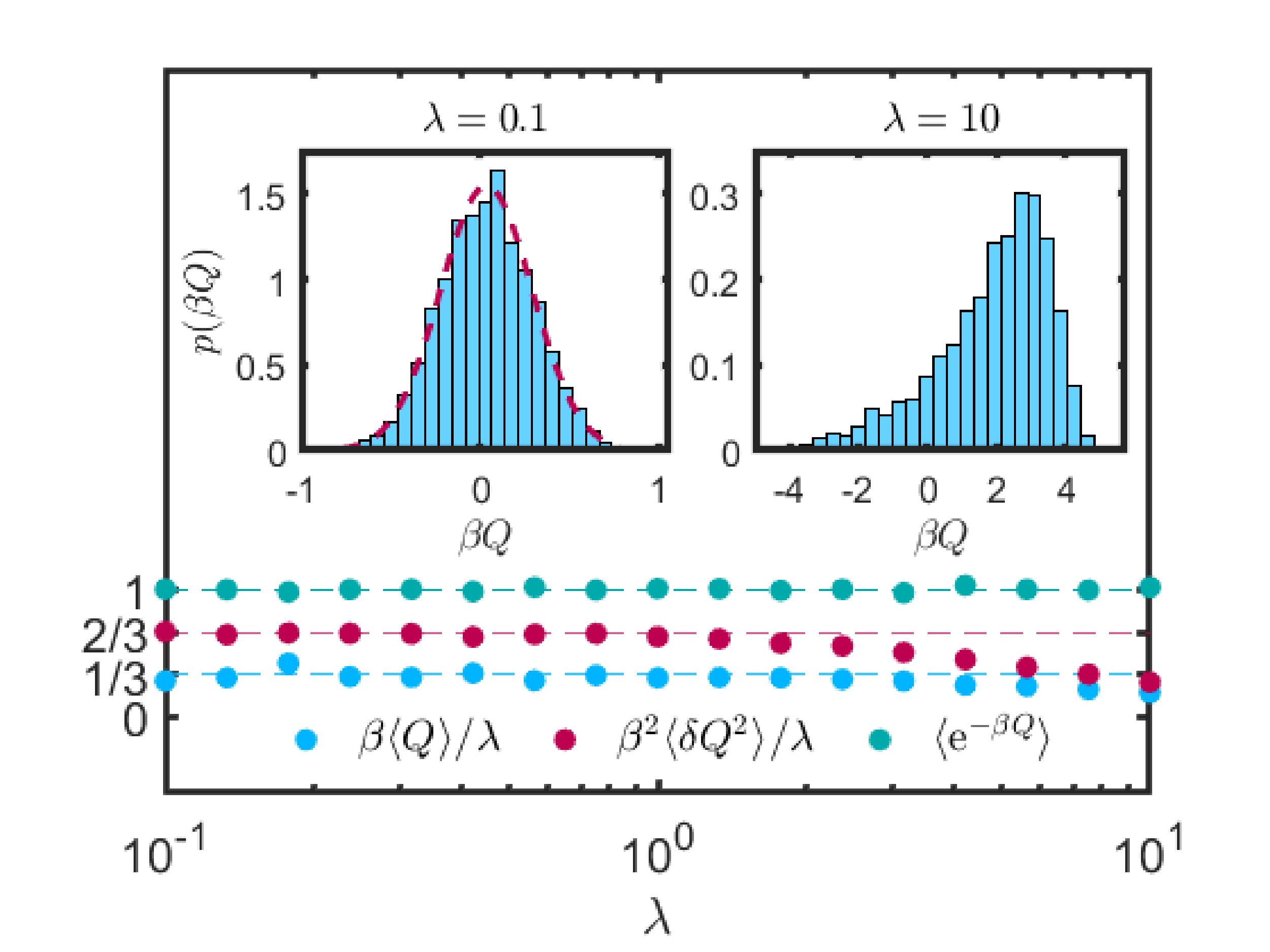}
    \caption{Heat statistics for the single-electron box. Stochastic simulations confirm that for $\lambda < 1$ the average heat and its fluctuations are accurately estimated by the adiabatic-approximation results from Eq.~\eqref{eq:NIS_FDR} and that above $\lambda \sim 1$ the approximation rapidly breaks down. The simulations also show that the integral fluctuation relation holds at all driving rates. The inset on the left shows that in the adiabatic-driving regime the energy dissipation becomes normally distributed as seen from the histogram and the Gaussian fit from Eq.~\eqref{eq:Gaussian}, displayed here with a dashed red line. Under faster driving, however, the distribution becomes negatively skewed, as can be seen from the histogram in the inset on the right.}
    \label{fig:histograms}
\end{figure}
\begin{align}
    \langle\delta Q^2\rangle = 2 k_\mathrm{B}T \langle Q\rangle = \frac{2}{3} \lambda (k_\mathrm{B}T)^2. \label{eq:NIS_FDR}
\end{align}
From this, we see that in the limit of adiabatic driving the mean and fluctuations vanish together as dissipation becomes suppressed even for individual ramps of the energy bias. 
To evaluate the validity of the adiabatic approximation, we carry out a comparison between numerically exact results and the adiabatic approximation. 
As shown in Fig.~\ref{fig:NIS_variances}, these results are in agreement in the slow-driving regime $\lambda \ll 1$ before diverging in the vicinity of $\lambda \sim 1$. 
The adiabatic approximation rapidly breaks down beyond that, proving that the value of the dimensionless driving rate functions as a good predictor of accuracy for the adiabatic approximation.

\textit{Fluctuation relations.---}
Numerical simulations of the driven transitions in the single-electron box show that the dissipated energy becomes normally distributed 
\begin{align}
    p(Q) = \sqrt{\frac{\beta}{4\pi\langle Q\rangle}} \, \mathrm{e}^{-\beta(Q-\langle Q\rangle)^2/4\langle Q\rangle} \label{eq:Gaussian}
\end{align}
in the limit of adiabatic driving $\lambda\ll 1$ as can be seen from the results of Fig.~\ref{fig:histograms}. 
To understand why the distribution becomes Gaussian under slow driving, we note that heat is exchanged during transitions between state $n=0$ and state $n=1$ and back. 
These transitions are independent of one another if they do not overlap, and their typical duration is captured by the correlation function $g(t_1,t_2)$. 
In the adiabatic limit, it takes considerably less time for the correlation to decay than for the gate voltage to change significantly. 
Thus, we can consider the total heat a sum of independent random variables of bounded variance, and because the central limit theorem applies, the distribution becomes Gaussian. 
A similar result was also obtained in an earlier study on the energy dissipated in driven transitions across a normal metal-insulator-normal metal tunnel junction \cite{Averin2011}. 
With our single-electron box, we see that in the adiabatic regime $\lambda \ll 1$ there is a high probability, $(1/2) \mathrm{erfc}(\sqrt{\lambda/24}) \simeq 1/2 - 0.12\sqrt{\lambda}$ to be precise, for energy to be extracted from the environment rather than dissipated during a single ramp. 
Combined with the fluctuation-dissipation relation, the Gaussian form of the distribution also implies the detailed fluctuation relation $p(-Q)/p(Q) = \mathrm{e}^{-\beta Q}$ and the associated integral fluctuation relation $\langle \mathrm{e}^{-\beta Q}\rangle = 1$. 
We also note that if the distribution of heat is Gaussian, the fluctuation relations imply the fluctuation-dissipation relation from Eq.~\eqref{eq:FluctuationDissipationRelation}, which can be verified with numerical simulations as seen in Fig.~\ref{fig:histograms}. 
However, it should be emphasized that these fluctuation relations for adiabatic driving are just a special case of a more general result and not unique to the Gaussian distribution as seen in the figure, which shows that the integral fluctuation relation is satisfied even in the fast-driving regime, where the adiabatic approximation breaks down and the distribution of heat becomes non-Gaussian. 
With our choice of driving protocol, both the forward and time-reversed trajectories begin and end in definite charge states, which combined with the fact that the protocol is odd under time reversal is sufficient to guarantee that the total entropy production is an involution with respect to time-reversal \cite{PelitiPigolotti}. 
As the entropy production is fully attributed to energy dissipation under our driving protocol, we can infer the fluctuation relations. 
While we have demonstrated these results for the superconducting single-electron box, similar results extend to other driving protocols with equivalent symmetry properties and boundary conditions as well as to other driven two-state setups with detailed balance.

\section*{Acknowledgements}

\noindent
This work was supported through Academy of Finland grant 312057 and by the Nokia Industrial Doctoral School in Quantum Technology.

\bibliography{main}

\end{document}